\documentclass[pra,aps,twocolumn,floatfix]{revtex4}
\usepackage{graphicx}
\usepackage{amsmath}

\begin{document}
\title[High fidelity state mapping performed in a V-type level structure]
{High fidelity state mapping performed in a V-type level structure via stimulated Raman transition}

\author{Grzegorz Chimczak}

\affiliation{Faculty of Physics, Adam Mickiewicz University, 61-614 Pozna\'n, Poland}

\email{chimczak@kielich.amu.edu.pl}
\pacs{03.67.Lx, 03.67.Hk, 42.50.Ct}

\begin{abstract}
It is proved that a qubit encoded in excited states of a V-type
quantum system cannot be perfectly transferred to the state of the cavity
field mode using a single rectangular laser pulse. 
This obstacle can be overcome by using a two-stage protocol, in which the
fidelity of a state-mapping operation can be increased to nearly one.
\end{abstract}

\maketitle
\section{Introduction}
The preparation and manipulations of photonic states by using an atom or
a quantum dot play an important role in quantum information
processing. Atoms, ions or quantum dots are
essential components of many optical quantum information processing
devices~\cite{kimble08:_quant_inter,northup2014quantum}, which have
been proposed~\cite{cirac97,enk98:_photon_chann_for_quant_commun,cabrillo99,bose,
duan_nature,duan:_effic,feng_entanglement,sun04:_atom_photon_entan_gener_and_didtr,
cho04, chou05:_measur_induc_entan_for_excit,
chimczak:_entanglement,chimczak:_entanglement_teleportation,
moehring07:_entan_of_singl_atom_quant,yin07:_multiat_and_reson_inter_schem,
wu07:_effec_schem_for_gener_clust,chimczak07:_improv,
beige2007repeat,busch08,ShiBiao08,chimczak09_nonmax,busch10,bastos12,
kyoseva2012coherent,yokoshi13}
or demonstrated~\cite{blinov04:_obser_of_entan_between_singl,
volz06:_obser_of_entan_of_singl, wilk07:_singl_atom_singl_photon_quant_inter,
boozerPRL07_map,choi08:_mappin_photon_entan_into_and,choi2010entanglement,
nolleke13:_efficient, gao2013teleportation, gao2012entanglement,reiserer14,pfaff14}
over past ten years. In such devices it is very useful to be able to transfer
qubit between the atomic state and the field state~\cite{parkins93:_synth_zeeman}.
Typically researchers perform a state-mapping operation using stimulated Raman
adiabatic passage (STIRAP)~\cite{kral07} because of the robustness of this technique
against different experimental imperfections. However, if the
state-mapping operation has to be really fast then STIRAP is not a proper
choice, since pulses should vary slow enough to fulfil the
adiabaticity criterion. If computational speed is very important
then the state-mapping operation via Raman transition should be based on
Rabi oscillations between these two states, in which qubit is encoded.
This method is much more demanding than STIRAP, but it is
probably that the near future technology will satisfy its all requirements.
In the paper~\cite{chimczak08:_fine} authors have discussed quantum
operations via Rabi oscillations performed in a three-level system in
the $\Lambda$-configuration with focus on improving fidelity.
The authors have shown that it is possible in this system to achieve high
enough fidelity to make these operations useful in future quantum computations,
i.e., the authors have shown that one can achieve the fidelity differing from
unity by $10^{-5}$, required by large quantum algorithms~\cite{preskill,steane99_doklad}.
Such high fidelities are a result of using fine tuning technique,
which prevents a reduction of the fidelity by the population of the third (auxiliary)
level. 

In this paper, we study the state-mapping operation performed in a 
V-type three-level system. The quantum interference manifested in the V-type system
leads to many important effects such as electromagnetically induced
transparency, quenching of spontaneous emission, lasing without inversion,
unexpected population inversion, quantum beats despite the incoherent pumping 
etc.~\cite{boller91,hakuta91,zhou96,zhou97,ficek00,ficek04,gong98,mompart00,
peng08,hegerfeldt93}, and therefore, this system is useful in quantum
information processing~\cite{turchette95,kojima09,cheng12,anton09,kim_agarwal99}.
The usefulness of V-type systems as memory elements is limited because of spontaneous
emission from excited states, however, if the time of a quantum
operation is much shorter than the decoherence time then it is possible 
to consider these systems as a candidate for qubit~\cite{imamoglu99,stievater01:_rabi,
feng03:_spin,feng03:_scheme,miranowicz:_gener,wang05_coherent}.
The aim of this paper is to show that the state-mapping operation in V-type
systems can be fast and the fidelity can be high enough to make
such systems useful in large quantum algorithms.

This paper is organized as follows. We begin in section 2 with a description of
the model. In section 3, we prove that it is impossible to transfer perfectly the qubit
encoded in excited atomic states to the state of a cavity field mode using a single
rectangular laser pulse. In section 4, we show that an approximate state mapping is
possible for long operation times only. In sections 5 and 6, we present the two-stage
state-mapping protocol that performs the transfer almost perfectly. Numerical results
(section 7) show that this protocol is fast and the fidelity satisfies the requirement
of large quantum algorithms. In sections 8 and 9, we investigate the influence of
field and atomic (respectively) damping on this protocol and we show that in some
special case the fidelity of the state mapping in the V-type system can be higher than
in $\Lambda$-type system.

\section{The model}
\begin{figure}[htbp]
  \centering
  \includegraphics[width=6.4cm]{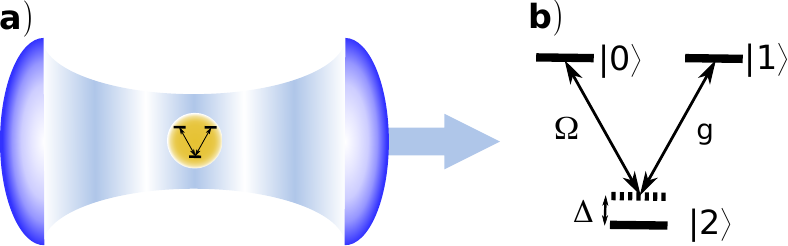}
  \caption{(a) Schematic setup for mapping a state of a V-type atom 
  (or an atom-like structure) onto the field state.
  (b)~V-type level configuration of the atom.}
  \label{fig:scheme}
\end{figure}

The state-mapping operation is performed using a device, which is formed of
an atom (or an atom-like structure) trapped inside a cavity. This atom
or atom-like structure plays the role of memory component and we
assume that it can be modeled by a three-level system in the V-configuration.
The setup and the level scheme are sketched in figure~\ref{fig:scheme}.
The quantum information is encoded in a superposition of two excited
levels $|0\rangle$ and $|1\rangle$. We can operate on this quantum
information using two transitions to the intermediate level $|2\rangle$,
which is the ground level. The first transition $|1\rangle\leftrightarrow|2\rangle$
is coupled to the cavity mode with a frequency $\omega_{\textrm{cav}}$
and coupling strength $g$. The second transition $|0\rangle\leftrightarrow|2\rangle$
is driven by a classical laser field with a frequency $\omega_{\textrm{L}}$
and coupling strength $\Omega$.
The classical laser field and the quantized cavity mode
are equally detuned from the corresponding transition frequencies by
$\Delta=(E_0-E_2)/\hbar-\omega_L$. 
The V-type level configuration cannot be considered as an ideal memory because,
contrary to the $\Lambda$-type system, a qubit is here encoded in a
superposition of two excited states, and thus, the time of storage of
quantum information is limited by the spontaneous emission.
Spontaneous emission rates from levels $|0\rangle$ and $|1\rangle$
we denote by $\gamma_0$ and $\gamma_1$, respectively.
However, new technology shows that it is possible to build
devices where the coherent coupling strength is much greater
than the spontaneous transition rate~\cite{englund07:_controlling,
englund10,englund12}. In~\cite{englund07:_controlling}
the coupling strength $g$ between quantum dot
and the microcavity mode is 80 times greater than spontaneous emission rate
$\gamma$ for the same transition. Therefore, the V-type level structure can be
considered as a short-term memory useful in quantum information processing.
Such large values of $g$ compared
to $\gamma$ are possible for microcavities since $g$ is inversely
proportional to the square of the cavity mode volume.
Unfortunately, the cavity decay rate $\kappa$ increases with decreasing
the cavity length, and therefore, $\kappa>g$ in~\cite{englund07:_controlling}.
Nevertheless, it is reasonable to assume that the finesse 
of microcavities will be improved in the future and 
the coherent coupling will dominate all dissipative rates.

The Hamiltonian that describes the interaction of the atom (or the quantum dot)
with the cavity field mode is given by
\begin{eqnarray}
  \label{eq:Hamiltonian0}
  H&=&-\Delta \sigma_{22}+(\Omega\sigma_{02}+g a \sigma_{12}
  + {\rm{h.c.}})\nonumber\\
  &&-i\kappa a^{\dagger}a-i\gamma_0\sigma_{00}
  -i\gamma_1\sigma_{11}\, ,
\end{eqnarray}
where $\sigma_{ij}=|i\rangle\langle j|$ denote the atomic flip operators
and $a$ denotes the annihilation operator of the cavity mode.

\section{Nonexistence of perfect mapping pulse for V systems}
Let us consider the case of the unitary evolution of the system, i.e.,
we assume that $\kappa$, $\gamma_0$ and $\gamma_1$ are equal to zero. 
The quantum information encoded in the atomic state
$\alpha|00\rangle+\beta|10\rangle$ has to be mapped perfectly and quickly
onto the cavity field state: $\alpha|00\rangle+\beta|01\rangle$.
Here, we have denoted a state of the system consisting of the atomic state 
$|j\rangle$ and the cavity field with $n$ photons by $|jn\rangle$.
So, we have to perform quantum operation defined by $|10\rangle\to|01\rangle$
and $|00\rangle\to|00\rangle$. Is it possible to
achieve this task using the evolution governed by the
Hamiltonian~(\ref{eq:Hamiltonian0})? Let us investigate this problem.
The evolution of the state $|10\rangle$ is given by
\begin{eqnarray}
  \label{eq:evolution10}
  e^{-i H t}|10\rangle&=&a(t)|10\rangle+b(t)|01\rangle+c(t)|21\rangle \, ,
\end{eqnarray}
where
\begin{eqnarray}
  \label{eq:solution0U7}
  a(t)&=&1-|g|^2/(|\Omega|^2+|g|^2)\,f(t)\, ,\nonumber\\
  b(t)&=&-\Omega\,g^{*}/(|\Omega|^2+|g|^2)\,f(t)\, ,\nonumber\\
  c(t)&=&-i g^{*}/\nu\,e^{i\,\Delta/2\,t}\sin{\nu\,t} \, ,
\end{eqnarray}
with $\nu=\sqrt{(\Delta/2)^2+|\Omega|^2+|g|^2}$ and
\begin{eqnarray}
  \label{eq:ft}
f(t)&=&1+e^{i\,\Delta/2\,t}(i\Delta/(2\nu) \sin{\nu\,t} 
  -\cos{\nu\,t})\, .
\end{eqnarray}
From~(\ref{eq:solution0U7}) one can see that the population
of the state $|10\rangle$ will be fully transferred to the state 
$|01\rangle$ if three conditions will be satisfied, i.e., $|g|=|\Omega|$,
$\sin{\nu\,t}=0$ and $f(t)=2$. Taking into account two first conditions
we can express the third one in the form
\begin{eqnarray}
  \label{eq:con0}
  \cos{(\Delta/2\,t_\pi)}\cos{\nu\,t_\pi}&=&-1 \, .
\end{eqnarray}
Equation~(\ref{eq:con0}) leads to a discrete set of detunings~\cite{chimczak08:_fine}
\begin{eqnarray}
  \label{eq:gen5}
  \Bigg(\frac{\Delta}{2\,|g|}\Bigg)^2&=&2\,\frac{\zeta^2}{2\zeta+1}  \, ,
\end{eqnarray}
where $\zeta=k/\theta$, $\theta=1, 3, 5,\dots$ is a natural odd number and
$k=0, 1, 2, 3,\dots$ is a non-negative integer. So, the population is
fully transferred from the state $|10\rangle$ to $|01\rangle$ if and
only if the value of detuning satisfies condition~(\ref{eq:gen5}) and 
the operation time $t_\pi$ is given by
\begin{eqnarray}
  \label{eq:tpi}
  t_\pi&=&2 k\pi/|\Delta|=(\theta +k)\pi/\nu \, .
\end{eqnarray}
If we assume that $\Omega=-g\exp(i\Phi)$ then it is seen from~(\ref{eq:evolution10})
and~(\ref{eq:solution0U7}) that such a perfect $\pi$ pulse is given by
\begin{eqnarray}
  \label{eq:Upi}
     U_\pi |10\rangle&\equiv&e^{-i H t_\pi}|10\rangle=e^{i\Phi}|01\rangle \, .
\end{eqnarray}

The state-mapping operation requires also that the population of the
state $|00\rangle$ has to remain unchanged. In the case of three-level $\Lambda$
systems the state $|00\rangle$ experiences no dynamics. 
Therefore, in such systems the perfect state-mapping operation (also defined by
$|10\rangle\to|01\rangle$ and $|00\rangle\to|00\rangle$) can be easily achieved.
The situation, however, is considerably more complicated for
three-level $\textrm{V}$ systems. In $\textrm{V}$ systems the time
evolution is given by
\begin{eqnarray}
\label{eq:200}
    e^{-i H t}|00\rangle&=&e^{i\Delta/2\,t}\Big[
  (2\nu'\cos{\nu'\,t}-i\Delta\sin{\nu'\,t})|00\rangle\nonumber\\
    &&-i 2\Omega^{*}\sin{\nu'\,t}|20\rangle\Big]/(2\nu') \, ,\\
\label{eq:201}    
    e^{-i H t}|20\rangle&=&e^{i\Delta/2\,t}\Big[
  (2\nu'\cos{\nu'\,t}+i\Delta\sin{\nu'\,t})|20\rangle\nonumber\\
    &&-i 2\Omega\sin{\nu'\,t}|00\rangle\Big]/(2\nu') \, ,    
\end{eqnarray}
where $\nu'=\sqrt{(\Delta/2)^2+|\Omega|^2}$. It is worth to mention
here that these above equations hold also for $\kappa\neq 0$.
During the evolution described by~(\ref{eq:200}) the population of
the intermediate state is given by
\begin{eqnarray}
  \label{eq:s22}
  \langle\sigma_{22}\rangle &=&
  \frac{4 |\Omega|^2}{\Delta^2+4 |\Omega|^2} \sin^{2}(\nu' t) \, .
\end{eqnarray}
The condition $|00\rangle\to|00\rangle$ is fulfilled up
to a phase factor if the operation time is given by
\begin{eqnarray}
  \label{eq:tpini}
t_\pi^{\prime}&=&l\,\pi/\nu'\, ,
\end{eqnarray}
where $l=1, 2, 3,\dots$ is a positive integer. Since the state-mapping
operation requires both conditions $|10\rangle\to|01\rangle$ and
$|00\rangle\to|00\rangle$, an operation time $t_{\rm{m}}$ has to
be equal to ~(\ref{eq:tpi}) and~(\ref{eq:tpini}), and the detuning
has to satisfy~(\ref{eq:gen5}). These three conditions lead to
a Diophantine equation
\begin{eqnarray}
  \label{eq:diophantine}
  k^2+(\theta +k)^2&=&2\,l^2 \, .
\end{eqnarray}
It follows from~(\ref{eq:diophantine}) that the numbers $k$ and $(\theta+k)$
should be both odd or both even. However, the numbers $k$ and $(\theta+k)$
will never be both odd or both even, because $\theta$ is always an odd number.
Thus,~(\ref{eq:diophantine}) has no solutions in natural numbers
$k$, $l$, $\theta$ with $\theta$ odd.

One can see that three-level $\textrm{V}$ systems have important
drawback. The fidelity of the state-mapping operation is always
reduced by the population of the intermediate level,
and therefore the perfect state mapping using single rectangular
laser pulse is impossible in three-level V-type systems.

Moreover, it seems likely that there is no prefect state mapping
consisting of more than one laser pulse. However, it is hard to prove 
it because there are infinitely many possible sequences of
laser pulses --- from two different pulses to series of very many
ultra-short pulses as in~\cite{kowalewska09}.

\section{Approximate state mapping}
We already know that there is no perfect state-mapping operation in
three-level $\textrm{V}$ systems, i.e., $t_\pi$ will be never equal to
$t_\pi^{\prime}$. However, $t_\pi$ can be very close to $t_\pi^{\prime}$
for some special numbers $k$, $l$, and $\theta$. In such cases it is
possible to perform an approximate state-mapping operation with
the operation time $t_{\rm{m}}\approx (t_\pi+t_\pi^{\prime})/2$.
Now we investigate if approximate state-mapping operations can satisfy
the fidelity requirement of large quantum algorithms. Of course, the fidelity
of approximate state-mapping operations is state dependent, and therefore,
we need the minimum fidelity taken over all possible input states in our
investigation. Since computational
speed is also very important, the times of these operations $t_{\rm{m}}$
should be not too long. We have calculated the minimal fidelity of state mapping
for all such $k$, $l$ and $\theta$, for which the operation time is shorter than
some fixed time limit $200 g^{-1}$. We have found that there are only fourteen different
state-mapping operations, which can satisfy the fidelity requirement of large
quantum algorithms $F\ge 1-10^{-5}$ and take less time than the chosen time limit
$200 g^{-1}$. We have only fourteen different values of the detuning, which we can
choose. The shortest approximate state-mapping operation lasts $109.5 g^{-1}$
and is determined by $(k, \theta, l)=(63, 17, 72)$. So, it is possible to
perform the approximate state mapping, but such an approximate state
mapping cannot be short and achieve very high fidelity at once.

\section{Almost perfect state mapping for V systems}
The state-mapping operation performed using a single rectangular laser pulse
in $\textrm{V}$-type systems cannot achieve fidelity equal to unity.
The population of the intermediate state $|2\rangle$ reduces the fidelity.
\begin{figure}[htbp]
  \centering
  \includegraphics[width=8cm]{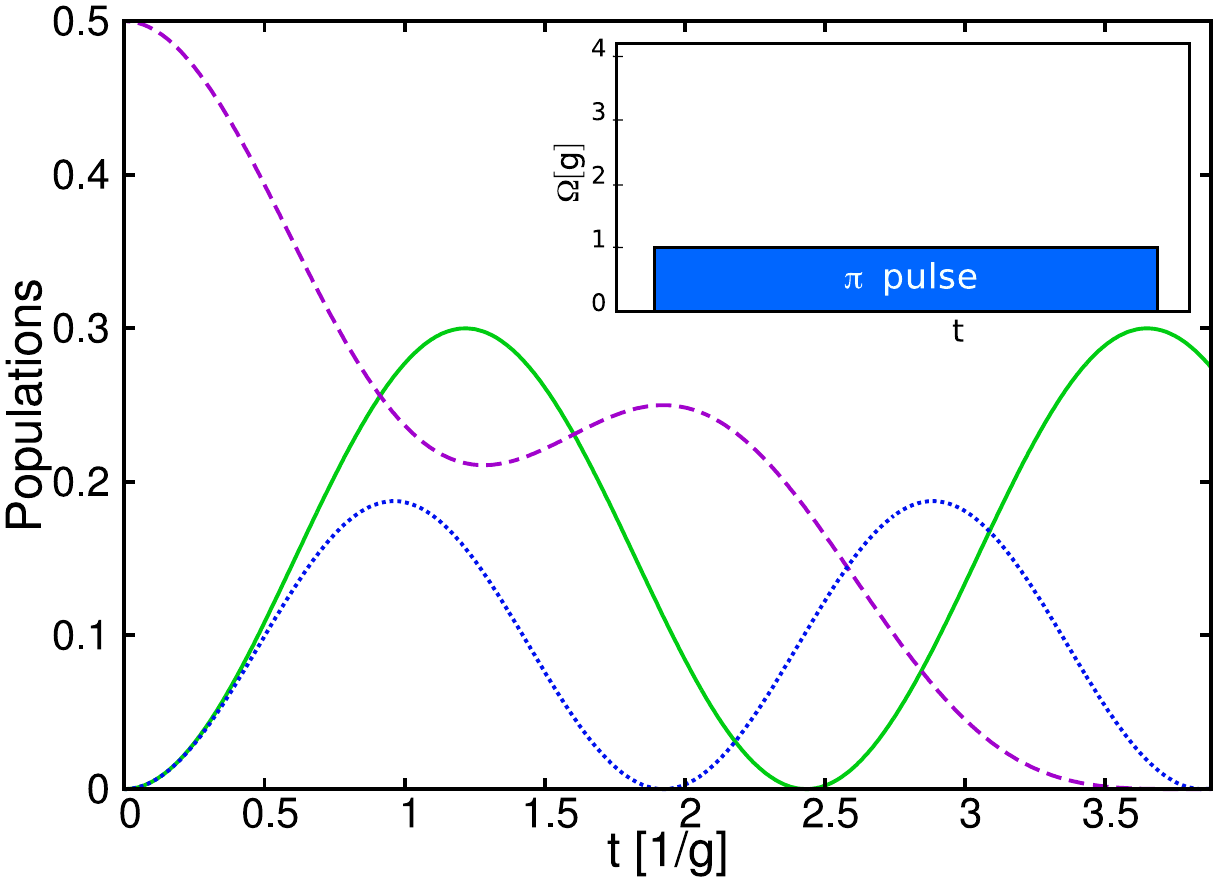}
  \caption{Populations of the states $|20\rangle$ (solid line), $|21\rangle$ (dotted line)
  and $|10\rangle$ (dashed line) during the $\pi$ pulse operation. A non-zero population
  of the state $|20\rangle$ at the end of the pulse reduces the fidelity. Inset: the $\pi$ pulse.
  Here we set $g=2\pi\times10$~MHz.}
  \label{fig:fig2}
\end{figure}
This situation is illustrated in figure~\ref{fig:fig2} --- the initial state
$(|00\rangle+|10\rangle)/\sqrt{2}$ cannot be perfectly transformed into 
$(|00\rangle+|01\rangle)/\sqrt{2}$ because of a non-zero population of the state
$|20\rangle$ at the end of the pulse.
The problem comes from the fact that the states $|10\rangle$ and $|00\rangle$
(which belong to orthogonal subspaces \{$|10\rangle$, $|01\rangle$, $|21\rangle$\}
and \{$|00\rangle$, $|20\rangle$\}) evolve with
noncommensurate frequencies $\nu=\sqrt{(\Delta/2)^2+2 |g|^2}$ and
$\nu'=\sqrt{(\Delta/2)^2+|g|^2}$ (for $|\Omega|=|g|$). The operations $|10\rangle\to|01\rangle$
and $|00\rangle\to|00\rangle$ require the $\pi$ pulse and the $2\pi$ pulse, respectively.
Since $\nu$ and $\nu'$ are noncommensurate, the duration times of these pulses 
are always different, and therefore, there is no rectangular pulse 
which can perform operations $|10\rangle\to|01\rangle$ and
$|00\rangle\to|00\rangle$ simultaneously.

We can omit this problem shifting the evolution in the subspace
\{$|00\rangle$, $|20\rangle$\} with respect to the evolution in the subspace
\{$|10\rangle$, $|01\rangle$, $|21\rangle$\} before performing the perfect $\pi$
pulse operation. By 'shifting the evolution' we mean that the state $|10\rangle$
experiences no dynamics while the state $|00\rangle$ is transformed into
such a special state $|\Phi_\eta\rangle=\eta_0 |00\rangle+\eta_1 |20\rangle\nonumber$
that $U_{\pi} |\Phi_\eta\rangle=|00\rangle$.
\begin{figure}[htbp]
  \centering
  \includegraphics[width=8cm]{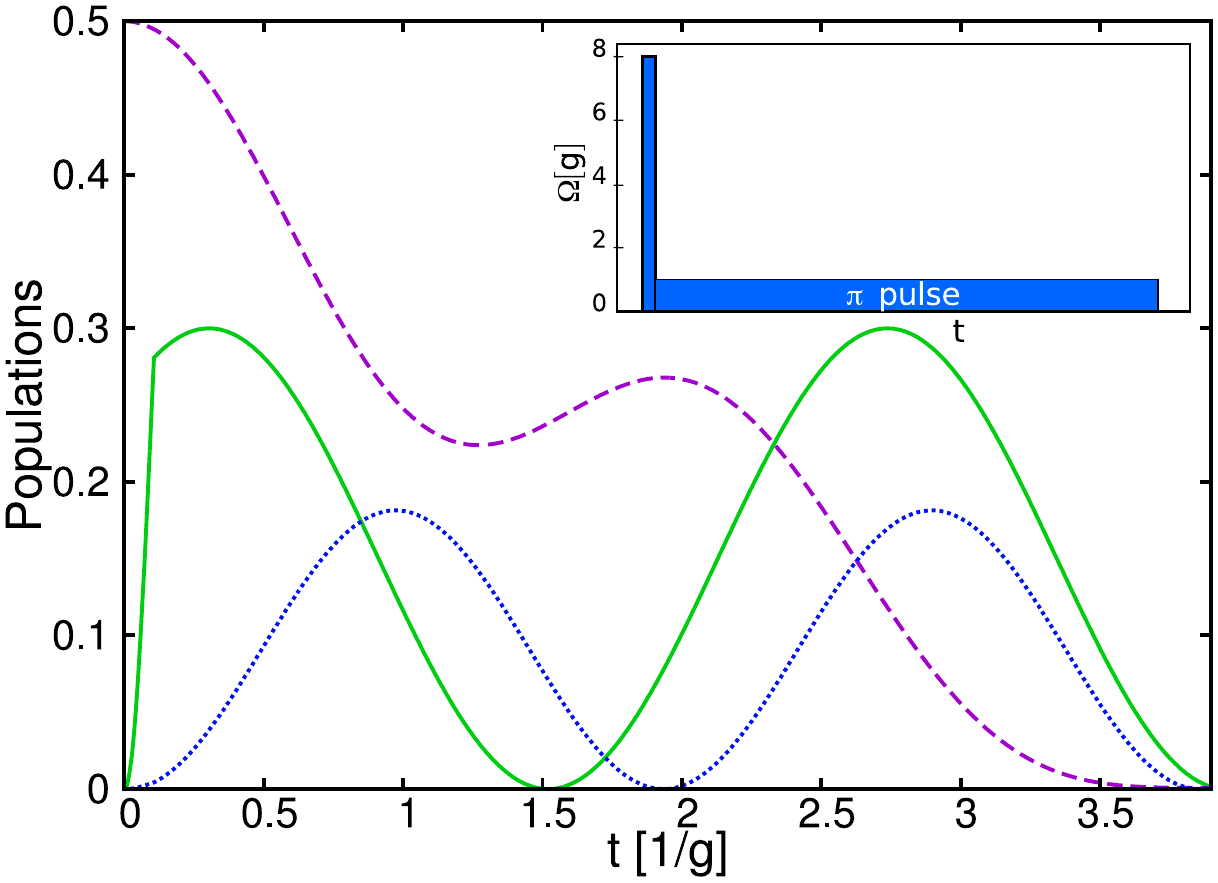}
  \caption{Populations of the states $|20\rangle$ (solid line), $|21\rangle$ (dotted line)
  and $|10\rangle$ (dashed line) during the two-stage state-mapping protocol. At the end
  these three populations are close to zero, and thus, the fidelity
  can be close to one. Inset: the pulses sequence.}
  \label{fig:fig3}
\end{figure}
The main idea of such a two-stage state-mapping protocol is
illustrated in figure~\ref{fig:fig3}. Thus, in the first stage of this protocol
we need an operation, which changes only the state
that belongs to the subspace \{$|00\rangle$, $|20\rangle$\} and leaves the state
belonging to the other subspace unchanged.

This needed operation is just the intense laser pulse operation. It is clearly seen 
from equations~(\ref{eq:evolution10}), (\ref{eq:solution0U7}) and~(\ref{eq:200}) 
that when the laser pulse is very intensive $|\Omega|\gg |g|$ then $a(t)\approx 1$,
and therefore, the state $|10\rangle$ experiences almost no dynamics while the population
of the state $|00\rangle$ oscillates with a very high frequency. In the following,
we will assume that $a(t)=1$ for $|\Omega|\gg |g|$. Since we need the fidelity greater
than $1-10^{-5}$, we have to estimate the error $\epsilon=1-|a(t)|^2$ introduced
by this approximation. From~(\ref{eq:solution0U7}) we see that
\begin{eqnarray}
  \label{eq:to01}
|a(t)|^2&>&1-\frac{|g|^2}{(|\Omega|^2+|g|^2)}\,2\,{\rm{Re}}({f(t_1)}) \, ,
\end{eqnarray}
where ${\rm{Re}}({f(t)})\leq 2$, and thus the probability that the system will be found in other
state than $|10\rangle$ is limited by
\begin{eqnarray}
  \label{eq:to02}
\epsilon&<&\frac{4\,|g|^2}{|\Omega|^2+|g|^2}\quad <\quad \Big(\frac{2\,|g|}{|\Omega|}\Big)^2 \, .
\end{eqnarray}
Therefore, $|\Omega/g|>633$ is necessary to get the error probability smaller than $10^{-5}$.

The effect of this operation is given by $U_1(t_1) |10\rangle=|10\rangle$ and
\begin{eqnarray}
  \label{eq:to03}
  U_1(t_1) |00\rangle&=&e^{i\Delta/2\,t_1}\big[
  (2\nu'_1\cos{\nu'_1\,t_1}-i\Delta\sin{\nu'_1\,t_1})|00\rangle\nonumber\\
  &&-i 2\Omega^{*}\sin{\nu'_1\,t_1} |20\rangle\big]/(2\nu'_1) \, ,
\end{eqnarray}
where $\nu_1'=\sqrt{(\Delta/2)^2+|\Omega|^2}$. 
From~(\ref{eq:to03}) it is seen that we can set arbitrary populations
of the states $|00\rangle$ and $|20\rangle$ for $|\Omega|\gg|\Delta|$.
It is also seen that we can easily give an arbitrary phase shift to the state
$|20\rangle$ with respect to the state $|00\rangle$ just by setting proper
argument of $\Omega=\Omega_1\exp(i\phi_\Omega)$, where $\Omega_1=|\Omega|$.
So we can always produce the state $|\Phi_\eta\rangle$ for large enough
$\Omega_1$.

We can calculate amplitudes $\eta_0$ and $\eta_1$ by premultiplying
$U_{\pi} |\Phi_\eta\rangle=|00\rangle$ by $U_{\pi}^{-1}=e^{i H t_{\pi}}$. 
We also assume that in the second stage
of the protocol, i.e., during the perfect $\pi$ pulse we change the intensity
and the phase of the laser field that $\Omega=-g\exp(i\Phi)$. In this way we get
\begin{eqnarray}
  \label{eq:to08}
  |\Phi_\eta\rangle&=&e^{-i\Delta/2\,t_{\pi}}\big[
  (2\nu_\pi'\cos{\nu_\pi'\,t_{\pi}}+i\Delta\sin{\nu_\pi'\,t_{\pi}})
  |00\rangle\nonumber\\
  &&-i e^{-i\Phi}2 g^{*}\sin{\nu_\pi'\,t_{\pi}}
  |20\rangle\big]/(2\nu_\pi') \,  ,
\end{eqnarray}
where $\nu'_{\pi}=\sqrt{(\Delta/2)^2+|g|^2}$.
A comparison of the moduli of the amplitudes~(\ref{eq:to03}) and~(\ref{eq:to08})
leads to
\begin{eqnarray}
  \label{eq:stage210b}
  t_1&=&\arccos{\Big(1-
  2\frac{|g|^2\nu_{1}^{\prime 2}}{\Omega_1^2\nu_{\pi}^{\prime 2}}
  (\sin{\nu'_{\pi}\,t_{\pi}})^2}\Big)/(2\nu'_1) \, .
\end{eqnarray}

In order to find the proper argument of $\Omega$ we write~(\ref{eq:to03})
and~(\ref{eq:to08}) 
in terms of the moduli and the arguments of $\eta_0$ and $\eta_1$
\begin{eqnarray}
  \label{eq:to09}
  U_1 |00\rangle&=&
  \big(|\eta_0| e^{i\theta_0}|00\rangle+|\eta_1| e^{i\theta_1}|20\rangle\big)
  e^{i\Delta/2\,t_1}\, ,\\
  |\Phi_\eta\rangle&=&
  \big(|\eta_0| e^{i\phi_0} |00\rangle+|\eta_1| e^{i\phi_1} |20\rangle\big)
  e^{-i\Delta/2\,t_{\pi}}\, ,
\end{eqnarray}  
where
\begin{eqnarray}
  \label{eq:to10}
  |\eta_1|&=&\frac{|g|}{\nu_\pi'}|\sin{\nu_\pi'\,t_{2\pi}}|\, ,\nonumber\\  
  |\eta_0|&=&\big(|g|^2\cos^2{\nu_{\pi}'\,t_{\pi}}+
  \Delta^2/4\big)^{1/2}/\nu'_{\pi}\, .
\end{eqnarray} 
The arguments of phase factors are given by
\begin{align}
  \label{eq:to11}
  \phi_1=\left\{
  \begin{array}{l}
  \frac{3}{2}\pi-\Phi-\phi_g\, ,\quad{\rm{if}}\quad\sin{\nu_\pi'\,t_{\pi}}>0\, ,\\
  \frac{1}{2}\pi-\Phi-\phi_g\, ,\quad{\rm{if}}\quad\sin{\nu_\pi'\,t_{\pi}}<0
  \end{array}\right.\nonumber
\end{align}
and
\begin{eqnarray}
  \label{eq:to12}
  \phi_0&=&
  \arctan\Big(\frac{\Delta}{2\nu_\pi'}\tan{\nu_\pi'\,t_{\pi}}\Big)+k'\pi\, ,\nonumber\\
  \theta_0&=&-\arctan{\Big(\frac{\Delta}{2\nu'_1}\tan{\nu'_1 t_1}\Big)}\, ,\nonumber\\
\theta_1&=&\frac{3}{2}\pi-\phi_\Omega\, ,
\end{eqnarray}   
where $\phi_g$ is the argument of $g$ and $k'$ is 0 or even for
$\cos{\nu_\pi'\,t_{\pi}}>0$ and odd for $\cos{\nu_\pi'\,t_{\pi}}<0$.

Now it is easy to check that if we chose such $\phi_\Omega$ that the condition 
$\phi_0-\phi_1+3\pi/2-\phi_\Omega-\theta_0=2\, l'\, \pi$ is fulfilled then
$U_1 |00\rangle=e^{i\Theta} |\Phi_\eta\rangle$, where
$\Theta=\Delta/2\,(t_1+t_\pi)+\theta_0-\phi_0$ and $l'$ is an integer.

\section{The state-mapping protocol}
The protocol that is able to achieve fidelity as close to unity as it
is needed consists of two stages: (A) the evolution-shift stage, and
(B) the $\pi$ pulse stage. Initially, the quantum system is prepared
in the state
\begin{eqnarray}
  \label{eq:pp01}
  |\psi_0\rangle&=&\alpha |10\rangle+\beta |00\rangle \, .
\end{eqnarray}
For simplicity, we assume here, and in the following, that $g$ is a real positive number.

\subsection{The evolution-shift stage}
The goal of this stage is to transform the state $|00\rangle$ into
$|\Phi_\eta\rangle$ without changing the state $|10\rangle$.
To this end, we turn the laser on for the time $t_1$, given by~(\ref{eq:stage210b}).
The laser has to be set in such a way that $\Omega=\Omega_1\exp(i\phi_\Omega)$ 
with $\phi_\Omega=\Delta/2\,(t_1+t_\pi)+m\,\pi$, where $m$ is 0 or even for
$\sin{\nu_\pi'\,t_{\pi}}>0$ and odd for $\sin{\nu_\pi'\,t_{\pi}}<0$.
The intensity of the laser light has to be great enough to satisfy the condition 
$\Omega_1\gg g$. This operation is described by $U_1 |10\rangle=|10\rangle$
and $U_1 |00\rangle=e^{i\Theta} |\Phi_\eta\rangle$, and therefore, at the end of
this stage, the system state is given by
\begin{eqnarray}
  \label{eq:pp02}
  |\psi_1\rangle&=&\alpha |10\rangle+\beta e^{i\Theta} |\Phi_\eta\rangle \, .
\end{eqnarray}

\subsection{The $\pi$ pulse stage}
In the second stage of this protocol we change the intensity of the laser
light to satisfy condition $\Omega=-g\exp(i\Phi)$ and we keep the laser on for
the time $t_\pi$. The $\pi$ pulse operation is described by
$U_\pi |10\rangle=e^{i\Phi} |01\rangle$ and
$U_\pi |\Phi_\eta\rangle=|00\rangle$. So, after this $\pi$ pulse operation the
system state is given by 
\begin{eqnarray}
  \label{eq:pp03}
  |\psi_f\rangle&=&\alpha e^{i\Phi} |01\rangle+\beta e^{i\Theta} |00\rangle \, .
\end{eqnarray}
If we set $\Phi=\Theta$ then the protocol ends up with the state
\begin{eqnarray}
  \label{eq:pp04}
  |\psi_f\rangle&=&\alpha |01\rangle
  +\beta |00\rangle\nonumber\\ 
  &=&|0\rangle_{\rm{dot}}\otimes(\alpha |1\rangle_{\rm{cav}}
  +\beta |0\rangle_{\rm{cav}}) \, .
\end{eqnarray}
The condition $\Phi=\Theta$ leads to $\Phi=\Delta/2\,(t_1+t_\pi)+\theta_0-\phi_0$.

\section{Validity of the rotating wave approximation}
Results obtained using~(\ref{eq:Hamiltonian0}) show that the fidelity of the 
state-mapping protocol tends to unity for large $|\Omega|$. However, we have to keep in 
mind that the Hamiltonian given by~(\ref{eq:Hamiltonian0}) describes the
quantum system composed of the V-type atom or atom-like structure and the
cavity in the rotating-wave approximation (RWA). This means that we cannot
set $|\Omega|$ too high because~(\ref{eq:Hamiltonian0})
will be unreliable~\cite{book_petruccione}.
The careful choice of $|\Omega|$ is especially important in the performing
of state mapping with a very high fidelity. Therefore, we have to take into account
in our considerations counter-rotating terms which are neglected in RWA.
The Hamiltonian without RWA is given by
\begin{eqnarray}
  \label{eq:rwa01}
  H&=&-\Delta \sigma_{22}+(\Omega\sigma_{02}+g a \sigma_{12}\nonumber\\
  &&+\Omega\sigma_{12} e^{i(\omega_{\rm{L}}+\omega_{\rm{cav}})t}
  +g a \sigma_{20} e^{-i(\omega_{\rm{L}}+\omega_{\rm{cav}})t}
  + {\rm{h.c.}})\nonumber\\
  &&-i\kappa a^{\dagger}a-i\gamma_0\sigma_{00}-i\gamma_1\sigma_{11}\, .
\end{eqnarray}
Here, we have assumed that the classical laser field and the quantized cavity
mode field are $\sigma^{-}$ and $\sigma^{+}$ polarized, respectively.
We can estimate the error introduced by the counter-rotating terms
using~(\ref{eq:rwa01}) and time-dependent perturbation theory. The quantum system,
which is initially prepared in the state $|10\rangle$ should remain in this state
after the first stage. Assuming that $|\Omega|,\,|\Delta|\ll\omega_{\rm{L}}+\omega_{\rm{cav}}$
and $|\Omega|\gg |g|$, the probability that it will be found in other state can
be roughly approximated by
\begin{eqnarray}
  \label{eq:rwa02}
\epsilon&=&\epsilon_1+\epsilon_2<\Big(\frac{2\,|g|}{|\Omega|}\Big)^2
	  +\Big(\frac{2\,|\Omega|}{\omega_{\rm{L}}+\omega_{\rm{cav}}}\Big)^2\, .
\end{eqnarray}
One can see that the first term of~(\ref{eq:rwa02}), which represents the error 
introduced by terms $g a \sigma_{12}+{\rm{h.c.}}$, is in agreement with~(\ref{eq:to02}).
The second term represents the error introduced by the counter-rotating terms.

Now it is easy to check that RWA is justified for atoms. For atoms, typically $g/2\pi$ 
is of order 10 MHz and $\omega_{\rm{L}}/2\pi\approx \omega_{\rm{cav}}/2\pi\approx 3.8\times 10^{8}$~MHz,
so $\epsilon_2$ is smaller than $10^{-6}$ up to $|\Omega|/g\approx 3\times 10^{4}$.
The situation is more complicated for quantum dots. We will consider the case of quantum
dots later.

\section{The numerical tests of the state-mapping protocol}
Let us see capabilities of the state-mapping protocol and check the derived formulas
using numerical computations. First, we examine the protocol for a great
value of the detuning $|\Delta|\gg g$. Here, and in the following,
we set $g=2\pi\times10$~MHz. For $(k, \theta)=(25, 1)$ and
$\Omega_1/g=100$ we obtain $\Delta/g=9.90148$, 
$(t_1, t_\pi)=(2.0\times 10^{-3}, 15.864)g^{-1}$,
$(\phi_\Omega, \Phi)=(0.01, 4.696)$ and $F>1-2\times 10^{-6}$.
It is seen that the total time of the protocol is much shorter
than the time of the shortest approximate state-mapping solution.
It is also worth to note that $t_1\ll t_\pi$, which means that the state-mapping
protocol for V-type systems is almost as fast as state mapping in $\Lambda$-type
systems. The fidelity is very close to one, so it is almost as high as
fidelity of state mapping in $\Lambda$-type systems. According to~(\ref{eq:to02})
we can increase the fidelity. The fidelity of the state-mapping protocol should
tend to unity as $\Omega$ becomes large. In order to check it we repeat these
calculation for $\Omega_1/g=1000$ and we obtain 
$(t_1, t_\pi)=(1.993\times 10^{-4}, 15.8642) g^{-1}$,
$(\phi_\Omega, \Phi)=(0.001, 4.6966)$ and $F>1-2\times 10^{-8}$.
As expected from~(\ref{eq:to02}), the error probability is proportional
to $(g/\Omega_1)^2$.

We can come to the same conclusions simulating the state-mapping protocol
for small values of $\Delta$. For example, $(k, \theta)=(1, 1)$ and
$\Omega_1/g=100$ lead to $\Delta/g=1.633$, 
$(t_1, t_\pi)=(8.5\times 10^{-3}, 3.847)g^{-1}$,
$(\phi_\Omega, \Phi)=(0.007, 4.319)$ and $F>1-7\times 10^{-5}$.
It is seen that the perfect $\pi$ pulse is faster for small
values of $\Delta$. It is also seen that the fidelity is smaller than
in case of large $\Delta$. This, however, is not a problem. We can always
increase the fidelity by increasing $\Omega_1/g$.
For $\Omega_1/g=1000$ we obtain $\Delta/g=1.633$, 
$(t_1, t_\pi)=(8.47\times 10^{-4}, 3.8476)g^{-1}$,
$(\phi_\Omega, \Phi)=(0.0007, 4.321)$ and $F>1-5\times 10^{-7}$.

\section{The effect of a non-zero $\kappa$ on the protocol}
The evolution of the state of real optical cavities is not unitary
because of absorption of photons in mirrors. In some of devices 
photons can also leak out of the cavity through a semitransparent mirror. Such a
photon leakage is very important when quantum information encoded in the photonic
state of the cavity has to be transferred to a distant quantum system.
We can take into account these photon losses assuming that cavity 
decay rate $\kappa$ is greater than zero. Let us now consider the
effect of non-zero $\kappa$ on operations needed by the state-mapping protocol,
i.e., $|10\rangle\to|01\rangle$ and $|00\rangle\to|00\rangle$.

As mentioned above, the evolution of the system prepared initially in 
the state $|00\rangle$ is independent of $\kappa$.
Hence, the time of the operation $|00\rangle\to|00\rangle$
is still given by~(\ref{eq:tpini}).
However, a non-zero $\kappa$ changes the evolution of the system prepared
initially in the state $|10\rangle$.
For small values of $\kappa$, we can find a good approximation to this
evolution applying the first order perturbation theory. In order to
write the expressions in a more compact form, we assume that
$|\Delta|\gg g\gg\kappa$. Then the expansion parameter can be
well approximated by $\eta=\kappa|\Delta|/(4 g^2)\ll 1$
and the evolution of the system is quite well described by
\begin{eqnarray}
  \label{eq:evol10k}
  e^{-i H t}|10\rangle&=&a_{\kappa}(t)|10\rangle
  +b_{\kappa}(t)|01\rangle+c_{\kappa}(t)|21\rangle \, ,
\end{eqnarray}
where
\begin{eqnarray}
  \label{eq:evol10k2}
  a_{\kappa}(t)&=&
    e^{-\kappa t/2}
    \big[
    e^{(i\omega_{+}-\kappa\xi_{+})t/2}
    \omega_{-}(1-i 2\eta\xi_{-})/(8\nu)-i\eta\epsilon\nonumber\\
    &&+e^{-(i\omega_{-}+\kappa\xi_{-})t/2}
    \omega_{+}(1+i 2\eta\xi_{+})/(8\nu)
    +1/2\big]\, ,\nonumber\\
  b_{\kappa}(t)&=&
    -e^{i\Phi}e^{-\kappa t/2} \big[
    e^{(i\omega_{+}-\kappa\xi_{+})t/2}
    \omega_{-}/(8\nu)-1/2\nonumber\\
    &&+e^{-(i\omega_{-}+\kappa\xi_{-})t/2}
    \omega_{+}/(8\nu)\big]\, ,\nonumber\\
  c_{\kappa}(t)&=&\frac{g}{2\nu}
    e^{(i\Delta-\kappa)t/2} \big(e^{-(i\nu+\kappa\xi_{-}/2)t}
    -e^{(i\nu-\kappa\xi_{+}/2)t}\big) \, .\nonumber\\
\end{eqnarray}
In~(\ref{eq:evol10k2}), $\omega_{\pm}=2\nu\pm\Delta$ and
$\epsilon={\rm{sign}}(\Delta)$ and $\xi_{\pm}=H(\pm\Delta)$,
where $H(t)$ denotes the Heaviside function.

From~(\ref{eq:evol10k2}),
it is easy to check that the population of the intermediate state
$|c_{\kappa}|^2$ is greater than zero for $t>0$. This means that the
state-mapping operation cannot be perfect for $\kappa>0$.
\begin{figure}[htbp]
  \centering
  \includegraphics[width=8cm]{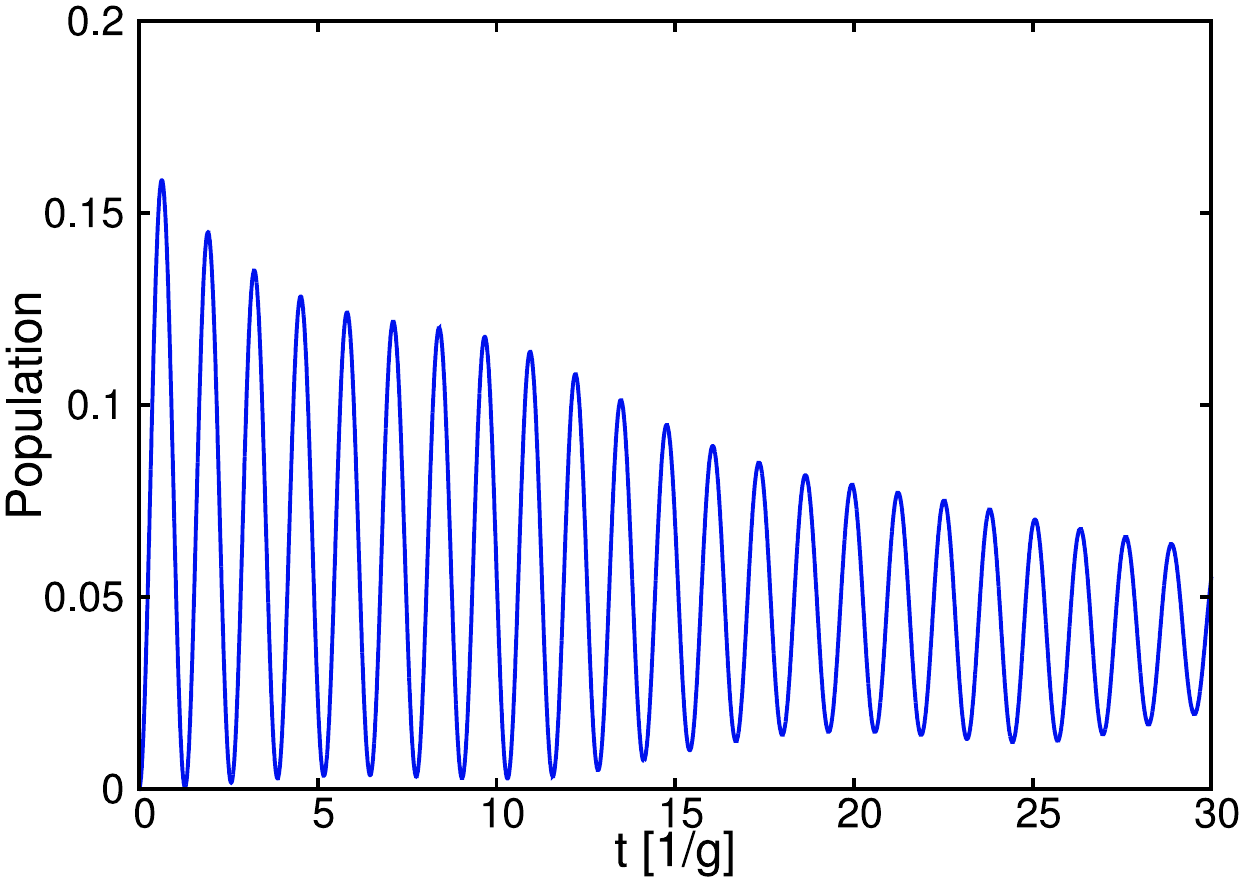}
  \caption{The population of state $|21\rangle$ computed
  numerically versus time for a non-zero cavity decay rate. This population
  reduces the fidelity of the state mapping. 
  The parameters are $(\Delta,\Omega,\kappa,\gamma_0,\gamma_1)/g=(4,1,0.1,0,0)$, where $g=2\pi\times10$~MHz.}
  \label{fig:fig4}
\end{figure}
In figure~\ref{fig:fig4} we plot the population of state $|21\rangle$
for quite large value of $\kappa$ to show that this population reaches
a local minimum periodically, but does not reach zero.
We can infer two consequences from this.
First, a very high fidelity is possible only for very small $\kappa t$.
Second, although we can only approximate the operation $|10\rangle\to|01\rangle$,
this approximation can be quite good if we use the oscillatory
behaviour of $|c_{\kappa}|^2$ to minimize it. It is interesting that
the local minima of $|c_{\kappa}|^2$ are almost independent of $\kappa$.
So, to a good approximation, this population takes minimum values at
\begin{eqnarray}
  \label{eq:Dka01}
  t=m'\pi/\nu \, ,
\end{eqnarray}
where $m'$ is a positive integer.

Second important difference in the operation $|10\rangle\to|01\rangle$
between cases when the damping is present and when is not is the time
of this operation. In order to get an approximated formula for the $\pi$
pulse, we need further approximations. Recalling that $|\Delta|\gg g\gg\kappa$, we drop all small terms and as a result we eliminate 
the state $|21\rangle$ from the evolution. Expressions for amplitudes obtained in this way are less precise, but much simpler
\begin{eqnarray}
  \label{eq:eli04}
  a_{\kappa}(t)&=&
    e^{-i\epsilon\delta t}e^{-\kappa t/2}\big(\cos{\delta t}+2\eta\sin{\delta t}\big)\, ,\nonumber\\
  b_{\kappa}(t)&=&i\epsilon e^{i\Phi} e^{-i\epsilon \delta  t} e^{-\kappa t/2} \sin{\delta t}\, ,\nonumber\\
  c_{\kappa}(t)&=&0\, ,
\end{eqnarray}
where $\delta=(2\nu-|\Delta|)/4\approx g^2/|\Delta|$.
The $\pi$ pulse requires $a_{\kappa}(t)=0$, and therefore the time
of this pulse is given by
\begin{eqnarray}
  \label{eq:tka02}
  t_{\pi}(\kappa)&=&
  \delta^{-1}\big[k''\pi-\arctan\big( (2\eta)^{-1}\big)\big]\, ,
\end{eqnarray}
where $k''=1,2,3,\dots$ Using the linear approximation, we can also express
$t_{\pi}(\kappa)$ in the form
\begin{eqnarray}
  \label{eq:tka12}
  t_{\pi}(\kappa)&=&\theta\pi/(2\delta)+2\eta\delta^{-1}
  =t_{\pi}(0)+2\eta\delta^{-1}\, ,
\end{eqnarray}
where $t_{\pi}(0)$ is given by~(\ref{eq:tpi}). This $\pi$
pulse is close to be perfect when the population of the intermediate
state takes minimum value at the end of this pulse. Therefore,
the time given by~(\ref{eq:tka12}) has to be equal to
that given by~(\ref{eq:Dka01}). This is possible only for
the fine tuned values of the detuning
\begin{eqnarray}
  \label{eq:Dka17}
  |\Delta(\kappa)|&=&|\Delta(0)|\Big(1-\frac{2\eta}{\theta\pi}\Big) \, ,
\end{eqnarray}
where $\Delta(0)$ is given by~(\ref{eq:gen5}).

The $\pi$ pulse operation for non-zero $\kappa$ can be approximated by
$U_\pi |10\rangle=e^{i(\Phi-\epsilon 2 \eta)} e^{-\kappa t_\pi/2} |01\rangle$ and
$U_\pi |\Phi_\eta\rangle=|00\rangle$. Therefore, after the second stage of
the protocol the unnormalized state of the system is given by
\begin{eqnarray}
  \label{eq:Dka18}
  |\widetilde{\psi}_f\rangle&=&
  \alpha e^{i(\Phi-\epsilon 2 \eta)} e^{-\kappa t_\pi/2} |01\rangle
  +\beta e^{i\Theta} |00\rangle \, .
\end{eqnarray}
Now it is useful to set $\Phi$, which satisfies the condition $\Theta=\Phi-\epsilon 2\eta$.
This leads to
\begin{eqnarray}
  \label{eq:Dka19}
  \phi_\Omega&=&\Delta/2\,(t_1+t_\pi)+m\,\pi+\epsilon 2 \eta\, ,\nonumber\\
  \Phi&=&\Delta/2\,(t_1+t_\pi)+\theta_0-\phi_0+\epsilon 2 \eta\, .
\end{eqnarray}
Then at the end of the protocol we obtain
\begin{eqnarray}
  \label{eq:Dka20}
  |\psi_f\rangle&=&{\cal{N}}(\alpha e^{-\kappa t_\pi/2}|01\rangle
  +\beta |00\rangle)\, ,
\end{eqnarray}
where ${\cal{N}}=(|\alpha|^2 e^{-\kappa t_\pi}+|\beta|^2)^{-1/2}$ is the normalization
factor. Observe that for non-zero values of $\kappa$ the state mapping is not perfect
because of the damping factor $e^{-\kappa t_\pi/2}$. We can achieve high
fidelity for small enough values of $\kappa t_\pi$ only.

Using~(\ref{eq:tka12}), (\ref{eq:Dka17}), (\ref{eq:Dka19}) and~(\ref{eq:stage210b})
we find that the minimal fidelity exceeds the value $1-10^{-5}$ for 
$(\Delta,\Omega_1,\kappa)/g=(9.890\,55, 1000, 7\times 10^{-4})$, 
$(t_1, t_\pi)=(2\times 10^{-4}, 15.882)g^{-1}$ and
$(\phi_\Omega, \Phi)=(0.008, 4.696)$.
More considerable value of $\kappa$ we can set for small values
of $\Delta$, because then the $\pi$ pulse is faster and $\kappa t_\pi$
smaller. The formulas~(\ref{eq:tka12}), (\ref{eq:Dka17}) and~(\ref{eq:Dka19})
work properly only when $|\Delta|\gg g\gg\kappa$ and $\eta\ll 1$. However, 
we can use these formulas to calculate the initial starting point and use 
it in a numerical optimization. In this way we find that we can achieve 
$F>1-10^{-5}$ for $(\Delta,\Omega_1,\kappa)/g=(1.630, 1000, 2.6\times 10^{-3})$, 
$(t_1, t_\pi)=(8.48\times 10^{-4}, 3.85)g^{-1}$ and
$(\phi_\Omega, \Phi)=(-0.001, 4.32)$. We cannot set larger values of $\kappa$
because $\kappa t_\pi$ is to large to achieve $F>1-10^{-5}$.

\section{The influence of spontaneous emission from excited states on
the protocol}
So far, we have assumed that spontaneous emission decay rates $\gamma_0$ and
$\gamma_1$ are equal to zero. Let us now relax this assumption
and investigate the influence of $\gamma_0$ and $\gamma_1$ on
state-mapping operations. It seems obvious that the fidelity of
state mapping decreases with increasing $\gamma_0$ and $\gamma_1$.
Sometimes even a small value of the spontaneous decay rate can considerably
decrease the fidelity or the success probability~\cite{chimczak02:_effect}.
Thus, it may be surprising that non-zero spontaneous emission decay rates
can improve state-mapping operations. The reason is that field 
and atomic damping act in a sense in opposite directions.
\begin{figure}[htbp]
  \centering
  \includegraphics[width=8cm]{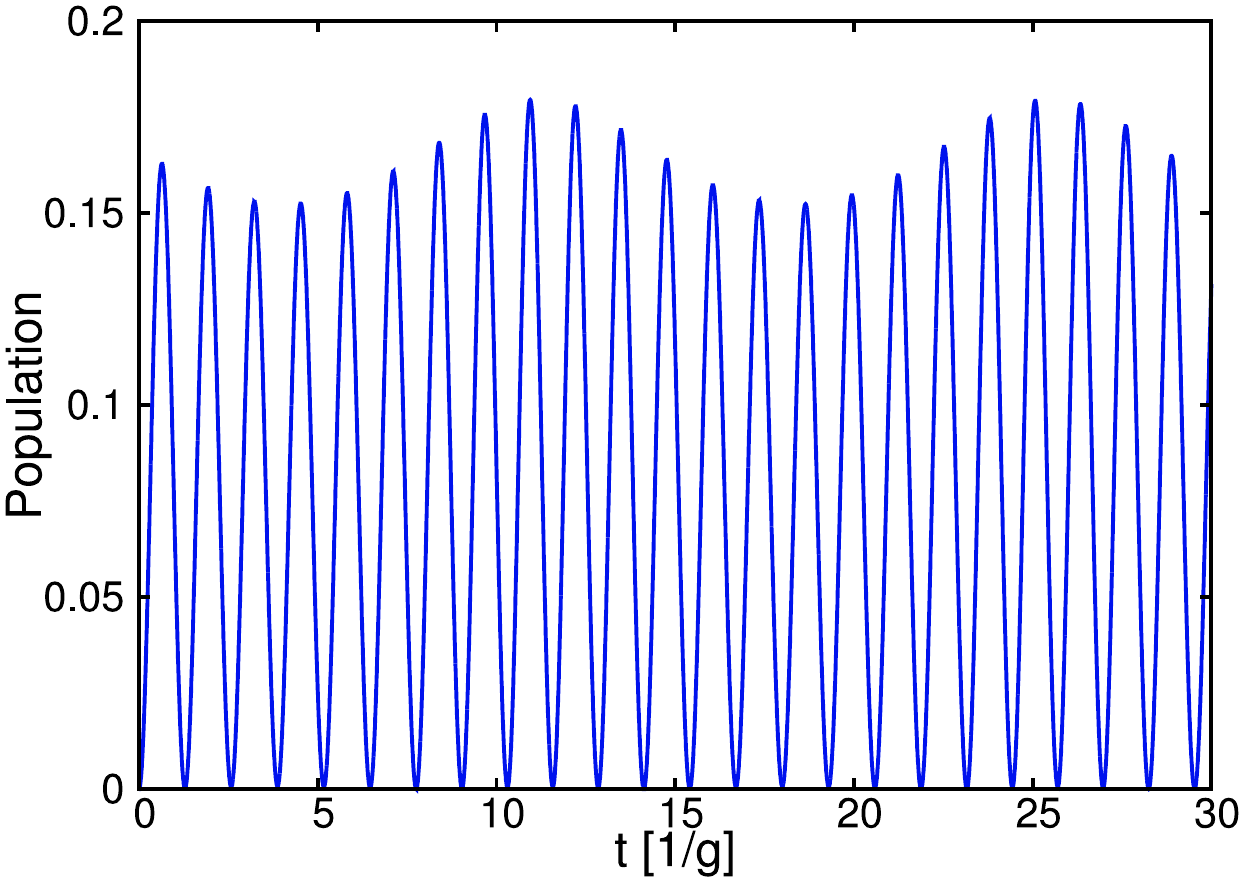}
  \caption{The population of the state $|21\rangle$
  as a function of time
  for the same parameters as in figure~\ref{fig:fig4}, but
  for ($\gamma_0,\gamma_1)/g=(0.05,0.05)$. For non-zero values of $\gamma_{0}$ 
  and $\gamma_{1}$ this unwanted population again approaches zero periodically.}
  \label{fig:fig5}
\end{figure}
Figure~\ref{fig:fig5} shows that the periodic behaviour of the system lost
due to non-zero $\kappa$ can be partially recovered by non-zero $\gamma_0$
and $\gamma_1$. The unwanted population of the state $|21\rangle$ again
approaches zero periodically. The same effect can be also observed in
the $\Lambda$-type system~\cite{chimczak08:_fine}.
This is one of rare examples of a decay process demonstrating its
usefulness in quantum-state engineering. However, it should be noted
that the atomic damping mechanism plays this constructive role only when
we are able to distinguish and reject unsuccessful cases of state-mapping operations,
where spontaneous emissions take place. Fortunately, in V-type systems
the quantum system is in the auxiliary level $|2\rangle$ after spontaneous
emission, and therefore, it is easy to check whether the spontaneous emission takes
place or not.

The atomic damping is responsible for one more surprise --- the V-type system
can be better than the $\Lambda$-type system for the state mapping.
The perfect $\pi$ pulse operation for non-zero $\kappa$, $\gamma_0$ and
$\gamma_1$ can be approximated by $U_\pi |10\rangle=e^{i(\Phi-\epsilon 2 \eta)}
e^{-(\kappa+\gamma_0+\gamma_1) t_\pi/2} |01\rangle$ and 
$U_\pi |\Phi_\eta\rangle=e^{-\Gamma t_\pi}|00\rangle$, where $\Gamma$
is an effective damping rate. If $|\Delta|\gg g\gg\kappa$ and $\eta\ll 1$
then $\Gamma\approx\gamma_0$. In this case, at the end of the protocol
the unnormalized state of the system is given by
\begin{eqnarray}
  \label{eq:gam01}
  |\widetilde{\psi}_f\rangle&=&\alpha e^{-(\kappa+\gamma_0+\gamma_1) t_\pi/2}|01\rangle
  +\beta e^{-\gamma_0 t_\pi}|00\rangle\, .
\end{eqnarray}
The normalized system state in the V-type system is given by
\begin{eqnarray}
  \label{eq:gam02}
  |\psi_f\rangle&=&{\cal{N'}}(\alpha e^{-(\kappa-\gamma_0+\gamma_1) t_\pi/2}|01\rangle
  +\beta |00\rangle)\, ,
\end{eqnarray}
while the normalized system state in the $\Lambda$-type system is almost independent of
$\gamma$ and can be well approximated by
\begin{eqnarray}
  \label{eq:gam03}
  |\psi_f\rangle&=&{\cal{N}}(\alpha e^{-\kappa t_\pi/2}|01\rangle
  +\beta |00\rangle)\, .
\end{eqnarray}
A comparison of~(\ref{eq:gam02}) with~(\ref{eq:gam03}) shows that the damping
factor in the V-type system can be closer to one than the damping factor in
the $\Lambda$-type system in the case of $\gamma_0>\gamma_1$ and large detunings.
Since the damping factor has significant influence on the fidelity for considerable
values of $\kappa$, the fidelity of the state mapping in V-type systems is higher 
in this case than in $\Lambda$-type systems. In~(\ref{eq:gam02}) 
${\cal{N'}}=(|\alpha|^2 e^{-(\kappa-\gamma_0+\gamma_1) t_\pi}+|\beta|^2)^{-1/2}$.

Now let us use atomic decay to increase $\kappa$ in the state-mapping
protocol without decreasing the fidelity. We want $\kappa$ as large as possible
because for real cavities it takes considerable values.
The state-mapping protocol,
for which the minimal fidelity exceeds $1-10^{-5}$, can be performed for 
$(\Delta,\Omega_1,\kappa,\gamma_0,\gamma_1)/g=(9.8568, 1000, 1.7\times 10^{-3},
1.7\times 10^{-3}, 7\times 10^{-4})$, $(t_1, t_\pi)=(1.98\times 10^{-4},
15.932)g^{-1}$ and $(\phi_\Omega, \Phi)=(-0.006, 4.696)$. Even larger value
of $\kappa$ can be set for small $\Delta$. Using numerical calculations, we
have found that $F>1-10^{-5}$ for
$(\Delta,\Omega_1,\kappa,\gamma_0,\gamma_1)/g=(1.619, 1000, 4.9\times 10^{-3},
9.2\times 10^{-3}, 2\times 10^{-3})$, $(t_1, t_\pi)=(8.44\times 10^{-4},
3.861)g^{-1}$ and $(\phi_\Omega, \Phi)=(-0.01, 4.32)$.

It is worth to note that it is possible to set $\kappa/g$ larger than
$4.9\times 10^{-3}$ with $F>1-10^{-5}$.
The main obstacle to achieve such high fidelity for large cavity decay rates
is the damping factor in~(\ref{eq:gam02}). We can overcome this
obstacle in the class of algorithms, in which the damping factor is compensated
for~\cite{ShiBiao08,chimczak09_nonmax}. In this way we can get $F>1-10^{-5}$
for $(\Delta,\Omega_1,\kappa,\gamma_0,\gamma_1)/g=(1.613, 1000, 1.8\times 10^{-2},
9\times 10^{-3}, 9\times 10^{-3})$, $(t_1, t_\pi)=(8.44\times 10^{-4},
3.87)g^{-1}$ and $(\phi_\Omega, \Phi)=(-0.009, 4.31)$.

\section{The state-mapping protocol in quantum dot systems}
A typical range of $g/2\pi$ in quantum dot-cavity systems is $8$ to
$38$~GHz~\cite{englund07:_controlling,winger08,englund10,englund12}, so
the coupling strength in quantum dot systems is three orders of magnitude
larger than in atom-cavity systems. Since the state-mapping protocol needs $|\Omega|\gg g$
in the first stage, counter-rotating terms become important and cannot be
neglected. One can check using~(\ref{eq:rwa02}) that for the excitonic
wavelength of~\cite{reinhard2012strongly} $\lambda=937.25$~nm 
($\omega_{\rm{L}}/2\pi\approx \omega_{\rm{cav}}/2\pi\approx 3.2\times 10^{5}$~GHz)
there is no such $|\Omega|$ that $\epsilon<10^{-5}$.
\begin{figure}[htbp]
  \centering
  \includegraphics[width=8cm]{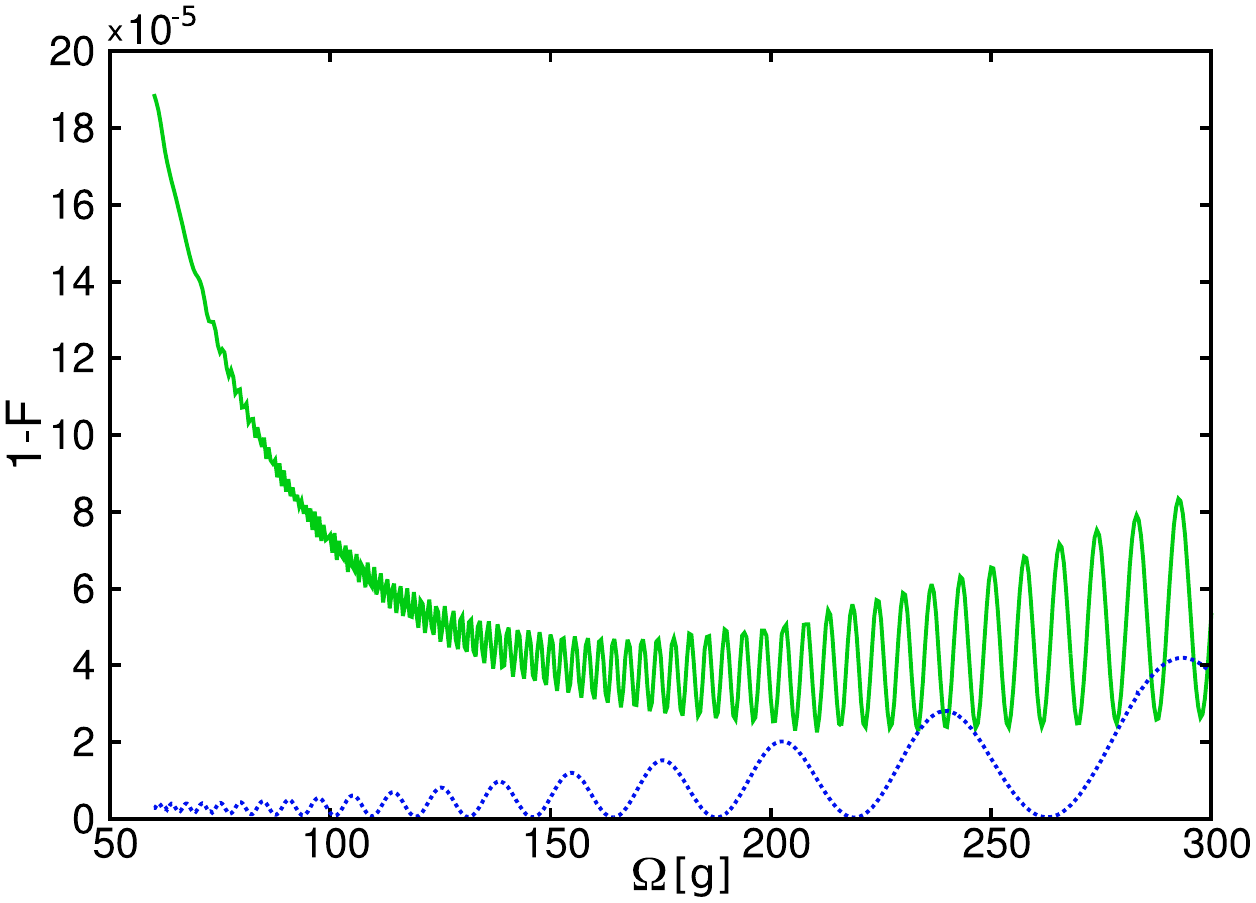}
  \caption{The effect of the counter-rotating terms on the fidelity for small value of $\Delta$
  (solid line) and for considerable value of $\Delta$ (dotted line).}
  \label{fig:fig6}
\end{figure}
It is seen in figure~\ref{fig:fig6}
that for $g=2\pi\times10$~GHz and small value of the detuning $\Delta/g=1.633$
($(k, \theta)=(1, 1)$) we obtain the fidelity, which does not satisfy the requirement of large
quantum algorithms, though it is still very high. 

Fortunately, $\epsilon_1$ can be reduced also by increasing $\Delta$. From~(\ref{eq:to01})
it is seen that $\epsilon_1$ is proportional to ${\rm{Re}}({f(t_1)})$, which tends to 0
as $\Delta\to\infty$. Figure~\ref{fig:fig6} shows that for large enough $\Delta$ it is
possible to perform the state-mapping protocol with $F\ge 1-10^{-5}$ by setting
moderate value of $\Omega_1$. We have obtained $F>1-2\times 10^{-6}$ for $g=2\pi\times10$~GHz,
$(k, \theta)=(60, 1)$ and $\Omega_1/g=62$ (which lead to $\Delta/g=15.4278$, 
$(t_1, t_\pi)=(2.08\times 10^{-3}, 24.435)g^{-1}$ and $(\phi_\Omega, \Phi)=(0.016, 4.705)$).
It is possible to satisfy the requirement of large quantum algorithms
even in the presence of field and atomic damping. The state mapping with the minimal fidelity
equal to $1-7\times 10^{-6}$ can be performed for $(\Delta,\Omega_1,\kappa,\gamma_0,\gamma_1)/g=
(15.4141, 62, 3.4\times 10^{-4}, 3.7\times 10^{-4}, 3.5\times 10^{-4})$, 
$(t_1, t_\pi)=(2.08\times 10^{-3}, 24.457)g^{-1}$ and $(\phi_\Omega, \Phi)=(0.021, 4.706)$.

It is worth to mention here that quantum optimal control theory~\cite{khaneja01time,schmidt11optimal}
makes it possible to manipulate spins very fast and with high fidelity in two level systems
beyond the RWA regime~\cite{scheuer14precise}. This is all what is needed in the first stage
of the state-mapping protocol performed in quantum dot-cavity systems. Therefore it is possible
that the presented results may be improved by using optimal control theory.

\section{Experimental feasibility of the protocol}
Finally, we shortly discuss the realizability of the state-mapping protocol
in a quantum system consisted of a quantum dot placed in a photonic crystal cavity,
like in~\cite{reinhard2012strongly}. A neutral exciton $X^{0}$ eigenstates naturally form a
three-level V-type system. Let us set experimentally achievable coupling strength
$g=2\pi\times10$~GHz and the exciton decay rate $\hbar\gamma=0.66\mu {\rm{eV}}$
($\gamma/g=0.016$)~\cite{reinhard2012strongly}. Let us also assume that the
damping factor is compensated for. Then we can obtain high fidelity
$F>1-1.3\times 10^{-4}$
for $(\Delta,\Omega_1,\kappa,\gamma_0,\gamma_1)/g=(1.599, 166, 3.2\times 10^{-2},
1.6\times 10^{-2}, 1.6\times 10^{-2})$, $(t_1, t_\pi)=(5.04\times 10^{-3},
3.888)g^{-1}$ and $(\phi_\Omega, \Phi)=(-0.0025, 4.30)$.
Note that the protocol time is short compared with $\gamma^{-1}$
of~\cite{reinhard2012strongly}. However, the protocol time is comparable to
$\gamma^{-1}_{\rm{deph}}$ of~\cite{reinhard2012strongly}, where $\gamma_{\rm{deph}}$
is the exciton pure dephasing rate. Moreover, as mentioned above, the value of the cavity
decay rate required by the protocol is demanding for present technology. In our numerical
calculations we have chosen value $\kappa/g=3.2\times 10^{-2}$, which is
40 times smaller than that of~\cite{reinhard2012strongly}.

\section{Conclusions}
I have shown that V-type quantum systems consisting of an atom or
atom-like structure and optical cavity have important drawback ---
quantum information stored in a superposition of two excited states
cannot be exactly mapped onto cavity mode state using a single rectangular
laser pulse. The fidelity of such a state mapping is always reduced by
the population of the intermediate ground state. 
However, I have found that there exists a two-stage state-mapping
protocol for V-type systems, which performs the state-mapping operation
almost perfectly, i.e., the fidelity tends to unity with increasing
the intensity of the laser light in the first stage of the protocol.
Since the first stage is ultra-short, this protocol is almost as fast
as state mapping performed in $\Lambda$-type quantum systems.
The protocol time is short compared with $\gamma^{-1}$ of~\cite{englund07:_controlling}. 
I have also investigated the influence of field and atomic damping
on this protocol. I have shown that the atomic decay can
be useful in the state-mapping protocol --- it can suppress unwanted
effects of the cavity decay. The atomic decay partially recovers the
periodic behaviour of the system and can make the damping factor close
to one. Surprisingly, in the limit of large detunings the state-mapping
protocol for V-type systems can achieve higher fidelity than the state 
mapping for $\Lambda$-type systems due to the atomic damping.

\begin{acknowledgments}
Fruitful discussion with Zbigniew Ficek is gratefully acknowledged.
This work was supported by Grant No. DEC-2011/03/B/ST2/01903 of the
Polish National Science Centre.
\end{acknowledgments}

\providecommand{\newblock}{}

\end{document}